\documentclass[11pt]{article}

\newif\ifwordcountmode
\ifdefined\wordcountmode
  \wordcountmodetrue
\else
  \wordcountmodefalse
\fi

\usepackage[a4paper,margin=1in]{geometry}
\usepackage[T1]{fontenc}
\usepackage[utf8]{inputenc}

\usepackage{microtype}
\usepackage{graphicx}
\usepackage{booktabs}
\usepackage{array}
\usepackage{multirow}
\usepackage{enumitem}
\usepackage{url}
\usepackage[hidelinks]{hyperref}
\usepackage{tabularx}
\usepackage{ragged2e}
\usepackage{array}
\usepackage{xurl}
\usepackage{tikz}
\usepackage{atbegshi}
\usepackage[british]{babel}
\usepackage{csquotes}
\usepackage[backend=biber,style=oscola,autocite=footnote,ibidstyle=uc]{biblatex}
\DeclareFieldFormat{url}{\url{#1}}
\AtEveryCitekey{\clearfield{note}}

\ifwordcountmode
  \renewcommand{\footnote}[2][]{}
\fi

\title{Where Trust Fails: \\ Mapping Location-Data Provenance Risks in Europe}
\author{%
Eduardo Brito$^{1,2}$ \quad Liina Kamm$^{1}$\\[0.35em]
$^{1}$Cybernetica AS, M\"aealuse 2/1, 12618 Tallinn, Estonia\\
$^{2}$University of Tartu, \"Ulikooli 18, 50090 Tartu, Estonia\\[0.35em]
\texttt{\{eduardo.brito, liina.kamm\}@cyber.ee}
}
\date{}

\AtBeginShipoutFirst{%
\begin{tikzpicture}[remember picture, overlay]
  \node[
    anchor=south,
    yshift=0.7cm,
    draw,
    inner sep=6pt,
    align=center,
    text width=0.7\paperwidth
  ] at (current page.south) {%
    \textbf{Preprint.} This work has been accepted for presentation at CPDP 2026, the 19th International Conference on Computers, Privacy and Data Protection.
  };
\end{tikzpicture}
}

\begin{document}
\maketitle
\thispagestyle{empty}

\vspace{-0.5em}
\begin{abstract}
European digital sovereignty and security increasingly depend on whether high-impact decisions can be grounded in location evidence that remains credible under adversarial pressure. This paper frames a cross-sector analysis as a \emph{location-data provenance} problem: whether there is contestable evidence about \emph{where and when} an asserted event occurred, \emph{who or what} produced the assertion, what a device or service reported as location, and which audit and retention guarantees apply. Trust-erosion patterns are observable across democratic processes and the information environment, trade and origin-sensitive supply chains, finance and illicit shipping flows, critical infrastructure and mobility, and harms targeting individuals' private and social domains. In these patterns we see a recurring asymmetry in which locality, presence, routing, or jurisdiction can be asserted cheaply while institutions and affected parties face costly reconstruction when disputes arise.

To make this challenge actionable, this paper introduces a compact risk taxonomy that decomposes provenance failures into integrity axes and recurring failure modes, and derives design expectations for next-generation digital trust infrastructure centered on \emph{contestability under dispute}, while remaining privacy- and rights-compatible. It argues for treating location as a digital primitive that should be represented as evidence-bearing claims rather than self-asserted coordinates, and positions proof-of-location (PoL) mechanisms as a candidate capability layer for producing verifiable presence claims under explicit threat and privacy assumptions. The outcome is a sector-neutral foundation for future architectural work on a next-generation digital trust infrastructure for Europe.

\end{abstract}


\vspace{0.5em}

\section{Introduction} \label{sec:intro}
Europe's digital sovereignty is increasingly tested in situations where regulatory, security, and governance decisions depend on the credibility of location evidence under adversarial conditions. Across digital platforms, supply chains, finance, and critical infrastructure, malicious actors can fabricate persuasive narratives and documentation at minimal cost, while verification processes remain slow, fragmented, and frequently dependent on a narrow set of intermediaries. In response, EU policy has progressively embedded evidence-grade governance as an operational expectation, most notably through systemic-risk obligations under the Digital Services Act (DSA) and targeted Commission guidance on safeguarding electoral processes \autocite{reg2022dsa,ec2024dsa_elections_guidelines}.

At the core of this challenge lies the problem of location-data provenance: the conditions under which claims of presence, routing, or jurisdiction can be examined, challenged, and reconstructed when disputes arise. In contemporary information operations, territorial context and grounding are routinely obscured through proxy infrastructures, synthetic identities, coordinated dissemination strategies, and manipulated or AI-generated media. EU threat analyses and election-related risk assessments consistently describe these dynamics as persistent, scalable, and difficult to attribute \autocite{eeas2024fimi_report,edmo2024eu_elections_risk_assessment}. The result is a structural imbalance in which locality can be asserted cheaply, while institutions and affected stakeholders incur significant costs to establish what actually occurred.

Similar dynamics recur across sectors with very different governance logics. In trade and customs, origin-sensitive compliance depends on credible claims of provenance and routing, yet enforcement repeatedly encounters relabelling practices and documentary substitution \autocite{dir2021aluminium_foil,dir2025biodiesel_protection}. In finance and fraud, cross-border opacity enables misrepresentation and regulatory arbitrage to scale \autocite{esma2021wirecard_peer_review,esma2021wirecard_td_letter}. In safety-of-life environments, positioning interference and spoofing degrade situational awareness and expose critical infrastructure and mobility services to hybrid risks \autocite{easa2024gnss_bulletin}. At the individual level, targeted impersonation, location spoofing, and deepfake media increasingly weaponise contextual ambiguity to impose reputational, professional, and psychological harm \autocite{ajder2019_state_of_deepfakes}.

This paper proceeds in four steps. Section~\ref{sec:sectors} grounds the diagnosis in European sectors where failures of location-data provenance already produce observable trust erosion. Section~\ref{sec:risks} introduces a compact taxonomy of recurring failure modes. Section~\ref{sec:implications} derives design expectations for next-generation digital trust infrastructure capable of supporting contestability and cross-border interoperability without defaulting to pervasive surveillance. Finally, Section~\ref{sec:location-primitive} positions location as a digital primitive and introduces proof-of-location (PoL) as a candidate capability layer for evidence-grade presence claims under explicit threat and privacy assumptions.

\section{Critical sectors and observable trust erosion} \label{sec:sectors}
This paper focuses on sectors where the integrity of location-data provenance is strategically relevant for European sovereignty and security. The selected sectors satisfy three criteria. First, location-linked claims in these domains can trigger legal, regulatory, economic, safety, democratic, or individual-rights consequences. Second, these claims are exposed to adversarial manipulation, evidentiary opacity, or costly reconstruction across organisational and jurisdictional boundaries. Third, failures in these domains already appear in operational incidents, enforcement practice, threat reporting, or technical literature, making them suitable for deriving recurring provenance risks.

The selected sectors provide variation across the main legal-technical functions of location. Democratic processes and the information environment test claims about event location, local observation, and jurisdictional context. Trade, customs, and supply chains test claims about origin, routing, and territorial eligibility. Finance, sanctions, and fraud test claims about jurisdiction, control, service delivery, and movement through regulated spaces. Critical infrastructure, mobility, and crisis response test claims about position, telemetry, navigation, and physical deployment. Individual targeting and private-sphere harms test claims about personal presence, context, and attribution. Together, these domains cover the principal ways in which location becomes governance-relevant while preserving sector-specific assurance levels and legal responses. The same analytical method can be extended to further domains where location-linked claims structure rights, obligations, access, attribution, or institutional trust, such as environmental monitoring, humanitarian response, public procurement, insurance, border management, and smart-city governance.

The cross-sector analysis follows a common legal-technical method. For each sector, we identify: \textbf{(i)} the location-linked claim on which institutional action or social trust depends; \textbf{(ii)} the mechanism by which that claim can be fabricated, replayed, recontextualised, laundered, proxied, obscured, or made difficult to reconstruct; and \textbf{(iii)} the governance consequence when legitimate actors cannot verify or contest the claim in time. This allows the paper to abstract from sector-specific incidents to recurring location-data provenance failures. The analysis identifies stable failure mechanisms that inform the taxonomy and infrastructure requirements developed in the following sections.

\subsection{Democratic processes and the integrity of the information environment}
Digital platforms have become critical infrastructure for civic discourse, political mobilisation, and electoral legitimacy. The EU has responded with binding obligations for systemic risk assessment and mitigation for very large online platforms and search engines \autocite{reg2022dsa}, complemented by Commission guidelines explicitly addressing electoral-process risks, including manipulative amplification, coordinated inauthentic behaviour, and the use of generative AI content \autocite{ec2024dsa_elections_guidelines}. The accompanying regulation on the transparency and targeting of political advertising further reflects an EU-wide recognition that information operations exploit opacity, cross-border delivery, and weak attribution of origin and targeting \autocite{reg2024political_ads}.

Recent enforcement signals illustrate the operational reality of these risks. During the Romanian elections, the Commission stepped up DSA monitoring of TikTok and issued a retention order to freeze and preserve data linked to systemic risks to electoral processes and civic discourse \autocite{ec2024romania_monitoring}. This led the Commission to open formal enforcement proceedings under the DSA, assessing, on the one hand, TikTok’s compliance with election-integrity risk mitigation obligations, and, on the other, whether specific platform systems may facilitate coordinated manipulation \autocite{ec2024tiktok_romania_proceedings}. These episodes align with broader assessments of foreign information manipulation and interference, and pose a persistent threat to European democratic resilience \autocite{eeas2024fimi_report,edmo2024eu_elections_risk_assessment}.

For location-data provenance, the democratic integrity problem is not limited to whether a post is authentic, but whether claims about where an event happened, who observed it, and under which jurisdiction it falls can be credibly validated when contested. Armed conflict illustrates how quickly fabricated locality can contaminate the public sphere. The war in Ukraine has already seen deliberately weaponised synthetic media at scale, starting with the deepfake surrender video attributed to President Zelenskyy \autocite{euronews2022zelensky_deepfake}, and the systematisation of the use of automated bots, deepfakes, decontextualised media, hashtag manipulation, and targeted messaging to spread false narratives, amplify propaganda, and manipulate public opinion \autocite{rehan2025ai}. The Israel--Palestine war has been accompanied by large volumes of miscontextualised footage and manipulated content that frequently hinges on false claims about where images were captured or what they depict \autocite{ap2023israel_hamas_misinfo}. This erosion of trust is compounded by what scholars term the “liar's dividend,” a phenomenon in which the mere existence of synthetic content enables actors to dismiss genuine recordings as fake, undermining accountability and trust in evidence \autocite{schiff2025liar}. In such settings, location-data provenance becomes a governance requirement for democratic processes, as an effort to reduce the probability that fabricated locality will be treated as evidence, and it lowers the cost of contestation when dispute is inevitable. 

\paragraph{Location-data provenance implication.}
For democratic processes and the information environment, \textbf{(i)} the location-linked claim concerns where an event occurred, who observed it, and whether the asserted territorial or jurisdictional context is credible; \textbf{(ii)} the failure mechanism consists of fabricated locality, recontextualised media, synthetic accounts, proxy infrastructures, coordinated amplification, and weak attribution of origin; and \textbf{(iii)} the governance consequence is that false territorial grounding can acquire evidentiary and political force before platforms, authorities, journalists, or affected communities can reconstruct the claim under dispute.

\subsection{Trade, customs, and origin-sensitive supply chains}
The European internal market depends on credible claims of origin, routing, and compliance. When location-linked assertions can be falsified or laundered through intermediaries, actors can relabel provenance, circumvent enforcement thresholds, and substitute documentation at scale. This dynamic appears in trade defense and circumvention contexts, where authorities must reconstruct whether products truly originate from declared jurisdictions and whether routing reflects substantive transformation or paper compliance \autocite{dir2021aluminium_foil,dir2025biodiesel_protection}. It also appears in customs and VAT fraud patterns that exploit cross-border complexity, where enforcement becomes reactive and document-heavy because the evidentiary chain is not verifiable end-to-end \autocite{eppo2025calypso}.

Supply-chain fraud scandals in food, pharmaceuticals, and consumer goods make the same point with higher safety stakes. Operation OPSON XIII, supported by Europol and Interpol with OLAF-led actions, seized large volumes of counterfeit or substandard food and beverages. They recorded systematic abuse of labelling, documentation, and geographical indications to infiltrate EU markets \autocite{olaf2024opson13,opson13_report}. In pharmaceuticals, Operation Pangea XVII resulted in major seizures, arrests, and the takedown of thousands of online listings used to distribute illicit medicines, showing how easily counterfeit or unapproved products can be marketed cross-border when provenance and distribution claims are weak \autocite{interpol2025pangea17}. In parallel, EU border enforcement reports show that counterfeit goods remain a sustained phenomenon across borders, with significant volumes detained across Member States and shifting origin and logistics patterns \autocite{ec2025counterfeit_detentions}. For location-data provenance, these cases share a structural feature. Once compliance is tied to origin, routing, or jurisdictional eligibility, adversaries exploit gaps between documentary assertions and verifiable evidence, and enforcement is forced into slow reconstruction rather than rapid verification.

This pressure is expanding as the EU imposes stricter due diligence regimes tied to geolocation and plot-level evidence. The deforestation-free products regulation, for instance, requires geolocation of production plots and due diligence statements \autocite{reg2023eudr}, increasing both the value of trusted location evidence and the incentives to falsify upstream claims when monitoring is weak or contested. As compliance becomes data-driven, the core question shifts from whether documents exist to whether the evidence behind them can be validated across organisational and national boundaries under dispute.

\paragraph{Location-data provenance implication.}
For trade, customs, and origin-sensitive supply chains, \textbf{(i)} the location-linked claim concerns origin, routing, production location, transshipment, and territorial eligibility for market access or regulatory treatment; \textbf{(ii)} the failure mechanism consists of relabelling, documentary substitution, laundering through intermediaries, fragmented supply-chain records, and unverifiable routing narratives; and \textbf{(iii)} the governance consequence is that enforcement shifts from timely verification toward slow reconstruction, allowing non-compliant goods, counterfeit products, or origin-sensitive fraud to scale across borders.

\subsection{Finance, illicit flows, sanctions, and fraud}
European financial integrity is repeatedly undermined by provenance failures that allow actors to misrepresent where economic activity takes place, which jurisdiction applies, and who exercises control. In cross-border finance, illicit flows increasingly exploit jurisdictional fragmentation and asymmetric access to evidence: shell structures, online platforms, and payment intermediaries enable rapid scaling of fraud while dispersing the evidentiary trail across multiple countries. Recent Eurojust-coordinated actions against large cryptocurrency and online investment scams demonstrate how professionally engineered platforms can defraud victims across many Member States, with critical evidence distributed among foreign service providers and intermediaries, delaying attribution and enforcement \autocite{eurojust2024crypto_scam,eurojust2025crypto_fraud_100m}.

Sanctions enforcement makes the location dimension explicit. Territorial claims about port calls, routing, and service delivery are routinely manipulated to evade oversight, turning location data from an operational input into a contested evidentiary object. In the Baltic Sea, documented GNSS interference and AIS spoofing affect commercial shipping, including tankers broadcasting falsified tracks consistent with attempts to conceal visits to Russian ports under sanctions pressure \autocite{easa2024gnss_bulletin,gfw2023tankers_ais_spoofing}. In both finance and sanctions contexts, the structural problem is the same: when high-impact decisions depend on location-linked claims that cannot be promptly contested with credible evidence, enforcement becomes reactive, supervisory authority weakens, and illicit activity scales faster than investigative reconstruction.

\paragraph{Location-data provenance implication.}
For finance, illicit flows, sanctions, and fraud, \textbf{(i)} the location-linked claim concerns jurisdiction, service delivery, operational control, asset movement, port calls, and the territorial exposure of regulated activity; \textbf{(ii)} the failure mechanism consists of opaque intermediaries, shell structures, falsified routing or tracking data, fragmented evidentiary trails, and strategic use of cross-border complexity; and \textbf{(iii)} the governance consequence is weakened supervisory authority, delayed attribution, and regulatory arbitrage in contexts where enforcement depends on reconstructing where activity occurred and who controlled it.

\subsection{Critical infrastructure, mobility, and crisis and conflict response}
In safety-of-life sectors, provenance failures create immediate operational risk, and in crisis governance they create strategic risk. Aviation and maritime operations already face contested navigation and positioning signals, with documented interference and spoofing affecting situational awareness and safety margins \autocite{easa2024gnss_bulletin}. At the same time, the EU's resilience policy recognises that hybrid threats blend cyber, information, and physical disruption, increasing the need for credible, auditable evidence in crisis response and continuity planning \autocite{dir2022cer}.

Conflict settings amplify the need for credible location-linked evidence while simultaneously degrading the trustworthiness of available signals and content. Response depends on timely, jurisdictionally meaningful facts: where an incident occurs, where affected assets are located, where responders are deployed, and where public evidence was collected. When these claims become contestable with disproportionate efficiency, coordination weakens and accountability becomes harder to sustain. For location-data provenance, the operational lesson is that evidence systems must function under stress and adversarial conditions, not only under normal operations.

\paragraph{Location-data provenance implication.}
For critical infrastructure, mobility, and crisis and conflict response, \textbf{(i)} the location-linked claim concerns position, navigation, telemetry, routing, responder deployment, asset location, and incident geography; \textbf{(ii)} the failure mechanism consists of signal spoofing, jamming, telemetry manipulation, degraded situational awareness, and contested public evidence collected under stress; and \textbf{(iii)} the governance consequence is operational uncertainty in safety-critical settings, weaker crisis coordination, and reduced accountability when institutional decisions depend on timely and credible location-linked facts.

\subsection{Individual targeting, private-sphere harms, and legal contestation}
Location-data provenance failures also enable individualised targeting, where synthetic or recontextualised media fabricates presence, intent, or misconduct around specific individuals. Deepfakes are central to this threat model, particularly in the form of non-consensual sexual content and image-based abuse. By exploiting evidentiary ambiguity, these practices impose reputational, professional, and psychological harm on victims. Empirical audits indicate that such content constitutes a substantial portion of detected deepfake material, highlighting private-sphere violations as a primary harm vector \autocite{ajder2019_state_of_deepfakes,sandoval2024threat}. The EU explicitly recognises this as a criminal-law and victim-protection issue. Directive (EU)~2024/1385 addresses non-consensual production or manipulation of intimate material, including AI-enabled deepfake-like fabrication, when made public without consent \autocite{dir2024_1385_vawg}.

These dynamics directly impact the legal system, as targeted synthetic media and impersonation attacks transform disputes about truth into disputes about provenance, authenticity, and admissibility of evidence. A systematic review highlights risks of evidence falsification, trust erosion in audiovisual records, attribution difficulties, and procedural strain, partly because courts and investigators lack consistent standards for rapid authentication under adversarial conditions \autocite{sandoval2024threat}. Europol frames deepfakes as a law-enforcement challenge affecting evidentiary reliability and investigative practice \autocite{europol2022_facing_reality_deepfakes}, while EU policy analysis stresses that effective mitigation requires coordinated responses across creation, dissemination, and verification rather than purely technical fixes \autocite{unit2021tackling}. From a location-data provenance perspective, the private-sphere problem mirrors the core diagnosis. Adversaries exploit cheap fabrication and rapid distribution, while victims and institutions face slow and uncertain contestation pathways, often hinging on falsified context about where an event allegedly occurred and whether a person was truly present and acting as claimed.

\paragraph{Location-data provenance implication.}
For individual targeting, private-sphere harms, and legal contestation, \textbf{(i)} the location-linked claim concerns a person’s alleged presence, conduct, context, association, or participation in an event; \textbf{(ii)} the failure mechanism consists of deepfake media, impersonation, recontextualised artefacts, falsified situational context, and rapid distribution before verification; and \textbf{(iii)} the governance consequence is reputational, professional, psychological, and procedural harm, with victims and institutions forced to contest fabricated or ambiguous evidence under severe time and information asymmetries.

\subsection{Deriving risks from cross-sector patterns} 

Across these sectors, the recurring pattern is that high-impact decisions depend on location-linked claims, adversaries exploit cheap assertion and costly verification, and institutional oversight is forced into slow reconstruction under asymmetric access. The sectoral implications above provide the analytical bridge from domain-specific incidents to recurring provenance failures: location-linked claims concern presence, origin, routing, jurisdiction, control, event location, or personal context; failure mechanisms include fabrication, replay, proxying, laundering, signal manipulation, record opacity, and asymmetric access; and governance consequences include delayed verification, weakened enforcement, disputed responsibility, evidentiary uncertainty, and pressure toward excessive data collection. Based on this cross-sector analysis, the next section presents a risk taxonomy for location-data provenance failures and provides a stable basis for deriving infrastructure requirements that remain meaningful across domains while preserving proportionality, contestability, and rights compatibility.

\section{Risk taxonomy: location-data provenance failures} \label{sec:risks}

This paper defines location-data provenance as the evidence, metadata, and governance controls that make a location-linked assertion contestable under dispute. It concerns whether competent parties can evaluate where and when something occurred, who or what produced the assertion, under which constraints, and with which audit and retention guarantees.

The taxonomy developed in this section abstracts the cross-sector patterns identified in Section~\ref{sec:sectors} into integrity dimensions and recurring failure modes. It should be read as a diagnostic framework for contested location-linked claims. A single incident may violate several integrity dimensions at once: for example, origin laundering can involve a false territorial claim, substituted documents, obscured control, and limited auditability. The purpose of the taxonomy is to identify where contestability breaks down in the lifecycle of a location-linked claim and to provide a stable basis for deriving infrastructure capabilities in Section~\ref{sec:implications}.

\begin{figure}[t!]
    \centering
    \includegraphics[width=0.6\linewidth]{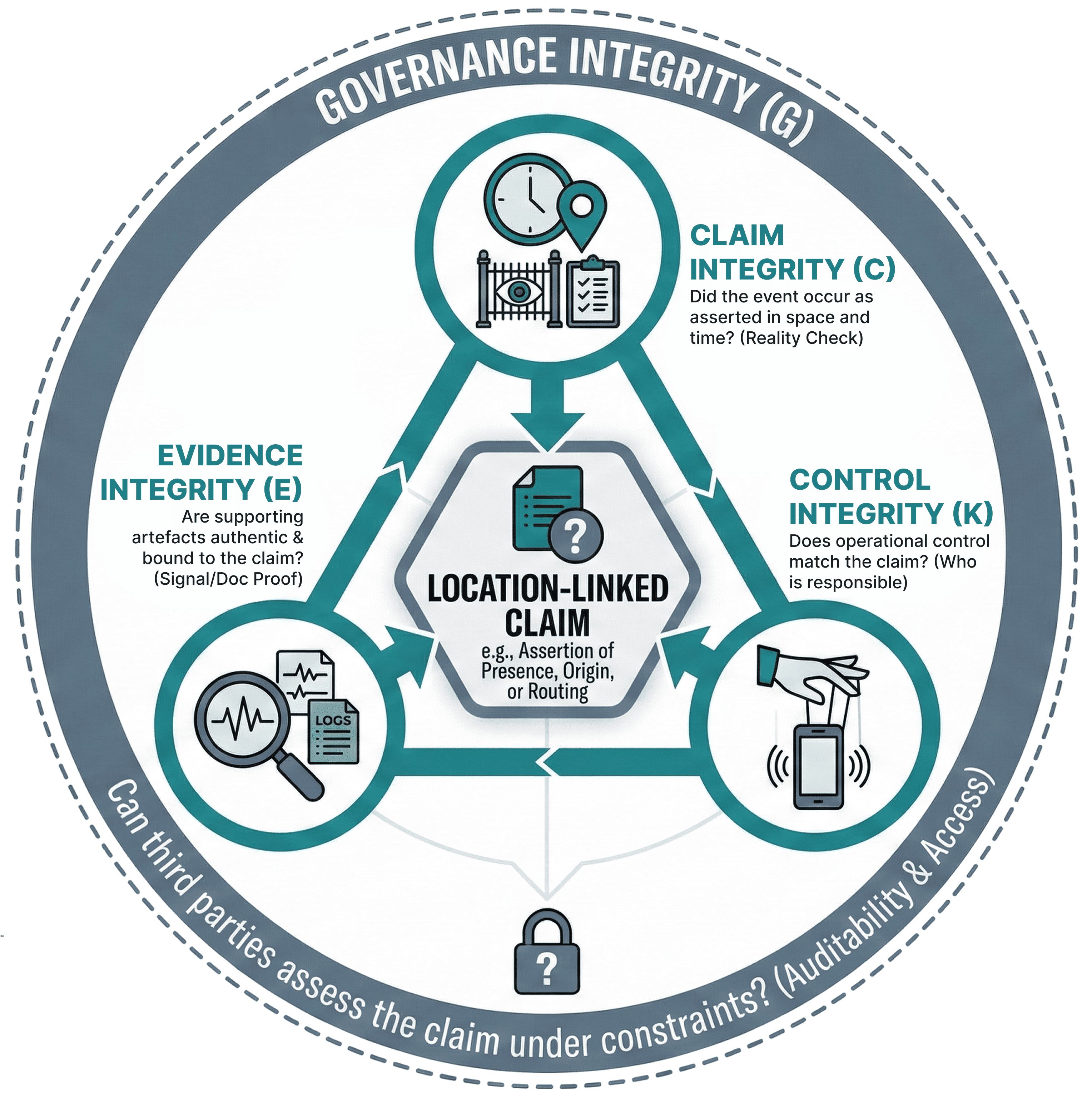}
    \caption{Location-data provenance integrity axes.}
    \label{fig:provenance_axes}
\end{figure}

In the EU context, this evidentiary framing matters because multiple instruments increasingly presuppose operational traceability for high-impact digital activity, including systemic-risk assessment and mitigation duties for very large online platforms and search engines (Digital Services Act \autocite{reg2022dsa}), with dedicated guidance for mitigating systemic risks affecting electoral processes (Commission guidelines C/2024/3014 \autocite{ec2024dsa_elections_guidelines}). It also includes transparency and due diligence obligations for political advertising services (Regulation (EU)~2024/900 \autocite{reg2024political_ads}), transparency obligations for certain AI-generated or manipulated content (AI Act \autocite{reg2024ai_act}), and complementary safeguards for media independence and pluralism in the digital internal market (European Media Freedom Act \autocite{reg2024emfa}). The Commission and the European Board for Digital Services have further endorsed the integration of the 2022 Code of Practice on Disinformation into the DSA framework as a Code of Conduct, reinforcing the compliance relevance of evidence-grade governance practices\autocite{ec2025code_disinfo_endorse,ec2025dsa_codes_conduct}. These instruments address different domains, but they share a common evidentiary orientation, that high-impact digital activity must increasingly be explainable, auditable, and contestable under institutional oversight. To make these expectations reusable across sectors and failure contexts, we treat provenance risk as an integrity problem and decompose it into four axes, illustrated in Figure~\ref{fig:provenance_axes}:

\begin{itemize}
    \item \textbf{Claim integrity (C):} whether the location-linked event occurred as asserted in space and time.
    \item \textbf{Evidence integrity (E):} whether supporting artefacts (signals, logs, media, documents) are authentic and bound to the claim.
    \item \textbf{Control integrity (K):} whether operational control matches what the claim implies, including delegation, proxying, or coercion.
    \item \textbf{Governance integrity (G):} whether third parties can assess the claim given asymmetric access, retention constraints, and privacy requirements.
\end{itemize}

\begin{figure}[t!]
    \centering
    \includegraphics[width=\linewidth]{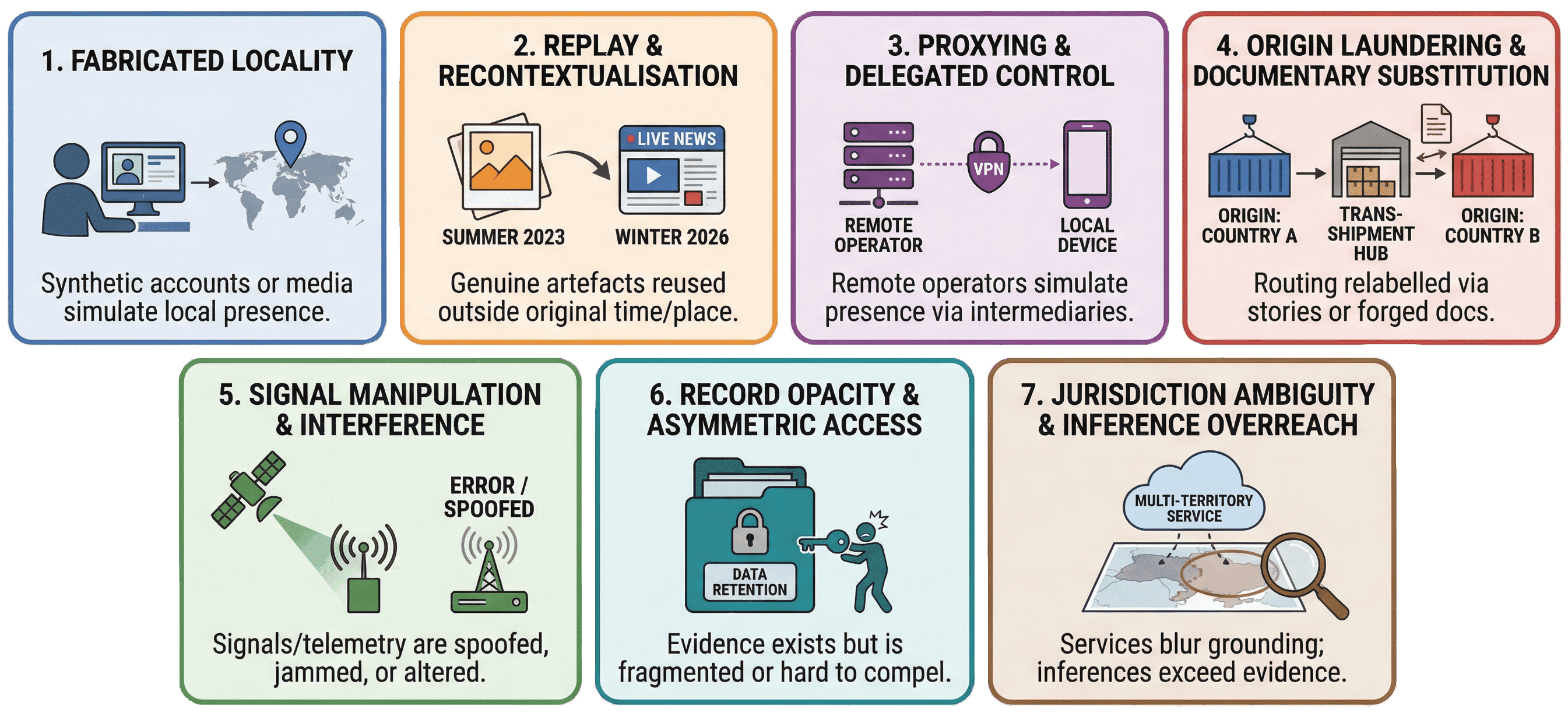}
    \caption{Illustration of the taxonomy of location-data provenance failures from Table~\ref{tab:taxonomy}.}
    \label{fig:provenance_layers}
\end{figure}

Table~\ref{tab:taxonomy} summarises recurring failure modes across the sectors in Section~\ref{sec:sectors} and annotates, for each mode, which axis or axes are primarily violated (C, E, K, G). Figure~\ref{fig:provenance_layers} illustrates how these failure modes arise across layers of location-data provenance, from physical signals to governance frameworks. A notable pattern in Table~\ref{tab:taxonomy} is that governance integrity (G) is implicated in most failure modes, reflecting that the bottleneck is often not the absence of signals, but the inability to audit and contest them across organisational boundaries. In many cases, evidence plausibly exists and may even be trustworthy, yet legitimate third parties cannot access, interpret, or audit it in time. The EU's implementation of the DSA shows a shift from ad hoc transparency toward structured evidence channels. Delegated Regulation (EU)~2025/2050 sets procedures and technical conditions for vetted researchers' access to Very Large Online Platforms (VLOPs) and Search Engines (VLOSEs) data, including a Commission-hosted portal and, where needed, secure processing environments \autocite{regdel2025_2050}, while Implementing Regulation (EU)~2024/607 establishes AGORA as the information-sharing system supporting supervision and enforcement \autocite{regimpl2024_607}. These instruments do not solve provenance on their own, but they demonstrate that auditability under constrained access and data protection requirements is now an institutional expectation. Enforcement practice also makes clear that evidence access is not optional. In December 2025, the Commission issued its first DSA non-compliance fine against X, citing failures including advertising repository transparency and researchers' access to public data \autocite{ec2025fine_x}. The effect for location-data provenance is direct, because when critical evidence is concentrated in a small number of intermediaries, sovereignty-relevant and public oversight depends on enforceable interfaces for scrutiny and contestation.

\begin{table}[t!]
\centering
\small
\renewcommand{\arraystretch}{1.12}
\setlength{\tabcolsep}{5.0pt}
\begin{tabularx}{\linewidth}{
    >{\RaggedRight\arraybackslash}p{0.22\linewidth}
    >{\RaggedRight\arraybackslash}p{0.10\linewidth}
    >{\RaggedRight\arraybackslash}X
    >{\RaggedRight\arraybackslash}X
}
\toprule
\textbf{Failure mode} & \textbf{Axes} & \textbf{What happens} & \textbf{Governance consequence} \\
\midrule
Fabricated locality & C,E,G
& Synthetic accounts or media simulate local observation or local presence.
& False territorial grounding becomes usable as evidence; rebuttal remains slow and intermediary-dependent. \\
\addlinespace

Replay and recontextualisation & C,E,G
& Genuine artefacts are reused outside their original time, place, or intent window.
& Disputes shift to narratives because freshness and context cannot be established reliably. \\
\addlinespace

Proxying and delegated control & K,G
& Remote operators simulate local presence via intermediaries, controlled devices, or compromised accounts.
& Attribution and jurisdiction become contestable; responsibility diffuses across layers. \\
\addlinespace

Origin laundering and documentary substitution & C,E,K,G
& Territorial origin or routing is relabelled through transhipment stories or forged documentation.
& Enforcement becomes reconstruction-heavy, enabling scale and plausible deniability. \\
\addlinespace

Signal manipulation and interference & E,C,K
& Location-bearing signals or telemetry are spoofed, jammed, or selectively altered.
& Operational trust degrades when safety and oversight depend on manipulable feeds. \\
\addlinespace

Record opacity and asymmetric access & K,G
& Evidence exists but is fragmented, privately held, inconsistently retained, or hard to compel in time.
& Compliance becomes intermediary-dependent, weakening auditability and contestability. \\
\addlinespace

Jurisdiction ambiguity and inference overreach & G,C
& Multi-territory services blur grounding; inferences exceed what available evidence supports.
& Over-trust invites exploitation; blunt controls incentivise over-collection and reduce openness. \\
\bottomrule
\end{tabularx}
\caption{Taxonomy of location-data provenance failures. ``Axes'' indicate which integrity dimensions (C, E, K, G) are primarily violated by each failure mode.}
\label{tab:taxonomy}
\end{table}

Two implications follow. First, because the failure modes combine claim, evidence, control, and governance failures, provenance cannot be reduced to a single signal source or a single attestation event. Robust governance requires evidence objects that remain interpretable across heterogeneous artefacts and organisational boundaries. Second, provenance must be compatible with privacy and proportionality. If institutions respond by defaulting to indiscriminate retention or pervasive surveillance, the result will be brittle compliance, contested legitimacy, and weak portability across sectors. Finally, new content provenance standards, driven by initiatives such as the C2PA, can strengthen transparency for media objects by structuring signed manifests of creation and modification actions \autocite{c2pa2024spec}. However, content provenance alone does not establish territorial credibility and is still vulnerable to manipulation of location signals, scene spoofing (e.g. AI generated imagery), and governance opacity. In this paper, it is treated as a complementary building block that can be composed with location-data provenance, rather than as a substitute for it.

The next section translates Table~\ref{tab:taxonomy} into design expectations for a next-generation digital trust infrastructure, focusing on contestability, privacy-preserving verification, cross-border interoperability, and auditability aligned with EU governance obligations.

\section{Implications for next-generation digital trust infrastructure}
\label{sec:implications}
The previous sections show that location-data provenance failures are rarely isolated technical faults. They are repeatable governance breakdowns that occur when high-impact claims become contestable, yet the institutional capacity to arbitrate those disputes remains dependent on opaque intermediaries, slow reconstruction, or overly broad data collection. The central design objective for a next-generation digital trust infrastructure targeting location data is therefore \emph{contestability under dispute}: the ability for legitimate actors to challenge location-linked claims and obtain a verifiable, rights-compatible resolution through structured evidence, defined procedures, and proportionate access mechanisms.

This section derives infrastructure implications from the taxonomy in Table~\ref{tab:taxonomy}. The derivation follows the rule: each principle or capability is justified by the role it plays in preventing one or more recurring provenance failures from reappearing under institutional or adversarial pressure. In this sense, the section moves from a diagnostic framework to a legal-technical requirements framework. The term \emph{infrastructure} is used broadly. It includes technical proof objects, verification interfaces, audit rights, retention rules, role separation, assurance semantics, and institutional access mechanisms that together make location-linked claims contestable across organisational and jurisdictional boundaries.

This objective aligns with the direction of EU governance for high-impact digital activity. The Digital Services Act frames systemic risks and mitigation as an accountability problem for very large platforms and search engines \autocite{reg2022dsa}, complemented by Commission guidance for electoral-process risks \autocite{ec2024dsa_elections_guidelines}. Regulation (EU)~2024/900 strengthens transparency and due diligence for political advertising \autocite{reg2024political_ads}, at the same time as the AI Act advances transparency expectations for certain AI-generated or manipulated outputs \autocite{reg2024ai_act}, and the European Media Freedom Act reinforces conditions for pluralistic and reliable media services \autocite{reg2024emfa}. The endorsed Code of Conduct on Disinformation further indicates that evidence-grade governance practices are treated as a practical benchmark for compliance \autocite{ec2025code_disinfo_endorse,ec2025dsa_codes_conduct}. These instruments operate in different domains, yet they share a common legal-technical orientation: high-impact digital activity increasingly requires traceability, auditability, and contestability under institutional oversight.

\subsection{Design principles for location-data provenance}

Table~\ref{tab:taxonomy} shows that location-data provenance failures recur because location-linked claims are increasingly used as evidence in governance contexts, while adversaries exploit the gap between cheap assertion and credible verification. The principles below define the normative and architectural constraints that a next-generation digital trust infrastructure must preserve for such claims to remain contestable under dispute. They are stated at the level of design principles rather than system components. Section~\ref{sec:requirements_summary} then translates them into minimum functional capabilities. The principles synthesize three sources of reasoning: the provenance failure modes identified in Table~\ref{tab:taxonomy}, EU governance expectations around accountability, traceability, and auditability, and technical literature on verifiable evidence, content provenance, secure positioning, and proof-of-location. A summary of the seven principles follows in Table~\ref{tab:principles_summary}.

\paragraph{Principle 1: Contestability by design.}
Location-linked claims must be challengeable through explicit verification procedures and evidentiary interfaces. This principle responds directly to fabricated locality, replay, recontextualisation, and record opacity, where disputes arise because content, logs, or documents circulate with weak means to check freshness, context scope, or binding to an original event. A structured contestation pathway allows a claim to function as evidence under dispute, because competent actors can test the claim against defined proof objects, metadata, and verification procedures. Contestability is therefore the foundational requirement for converting location from an asserted attribute into an assessable evidentiary claim.

\paragraph{Principle 2: Proportionality and rights compatibility.}
Because location is inherently sensitive, provenance mechanisms must support dispute resolution through selective, purpose-bound disclosure. This principle responds to governance integrity failures and jurisdictional ambiguity, where the demand for verifiability can create pressure toward broad retention or excessive inference. The infrastructure should make the minimum evidence needed for a specific claim available under a defined legal basis, assurance threshold, and access condition. Proportionality preserves legitimacy, reduces over-collection incentives, and makes cross-sector deployment compatible with privacy and fundamental-rights requirements.

\paragraph{Principle 3: Separation of roles.}
The repeated pattern of record opacity and asymmetric access shows that concentrating evidence generation, retention, verification, and adjudication in a single intermediary creates a structural vulnerability. Location-linked claims often traverse multiple actors, including platforms, brokers, service providers, infrastructure operators, and public authorities. A next-generation approach should support role separation such that evidence can be generated by one party, presented by another, verified by a third, and assessed under defined institutional procedures. This reduces conflicts of interest, increases auditability, and makes trust assessment possible even when a dominant intermediary is itself implicated in the disputed claim.

\paragraph{Principle 4: Accountability for delegation and proxying.}
Proxying and delegated control are central adversarial strategies in location-data provenance failures because they allow remote operators to simulate local presence while preserving plausible deniability. Provenance must therefore treat control context as first-class. When a location claim implies presence, origin, routing, or jurisdiction, the evidentiary question includes whether the claimant, device, organisation, or intermediary exercised the kind of operational control that the claim presupposes. Accountability for delegation allows verifiers to distinguish direct presence, mediated presence, compromised control, coercion, and remote operation, which is necessary for responsibility attribution under dispute.

\paragraph{Principle 5: Resilience to laundering, substitution, and manipulation.}
Origin laundering, documentary substitution, and signal manipulation show that adversaries can engineer evidentiary paths that survive superficial scrutiny. Provenance mechanisms must assume compromised sources, forged artefacts, intermediary laundering, and manipulated telemetry as realistic conditions. The infrastructure should therefore support corroboration across heterogeneous artefacts, detectability of substitution points, and explicit uncertainty semantics where evidence is incomplete or degraded. The objective is to increase the cost of deception, shorten time-to-resolution, and preserve institutional confidence when single evidence sources become unreliable.

\paragraph{Principle 6: Portability across borders and sectors.}
A location claim becomes governance-relevant precisely because territorial regimes differ. The same failure modes recur across elections, trade, crisis response, financial compliance, sanctions enforcement, and private-sphere harms. Provenance mechanisms must therefore be portable across jurisdictions, sectors, and institutional boundaries. Portability requires shared evidence semantics, declared assurance assumptions, and verification procedures that remain interpretable by different competent actors. It also preserves sector-specific thresholds, legal tests, and risk tolerances, so that common evidence objects can support differentiated governance decisions.

\paragraph{Principle 7: Governance-grade auditability.}
The taxonomy indicates that many failures persist because evidence cannot be examined by legitimate oversight actors in time. This is a governance integrity problem as much as a technical integrity problem. A next-generation digital trust infrastructure must therefore provide auditable interfaces, retention semantics, and access procedures that make location-linked claims assessable under appropriate safeguards. Governance-grade auditability ensures that contestability remains operational when critical evidence is held by private intermediaries, distributed across jurisdictions, or subject to data protection constraints.

\begin{table}[h!bt]
\centering
\small
\renewcommand{\arraystretch}{1.12}
\setlength{\tabcolsep}{5.5pt}
\begin{tabularx}{\linewidth}{
    >{\RaggedRight\arraybackslash}p{0.30\linewidth}
    >{\RaggedRight\arraybackslash}X
}
\toprule
\textbf{Design principle} & \textbf{Meaning for location-linked claims and evidence} \\
\midrule
Contestability by design
& Location-linked claims must be challengeable through explicit verification procedures and evidentiary interfaces that can resolve disputes about freshness, context, and binding to an event. \\
\addlinespace
Proportionality and rights compatibility
& Provenance should enable dispute resolution through selective, purpose-bound disclosure commensurate with the claim, legal basis, and assurance need. \\
\addlinespace
Separation of roles
& Evidence generation, presentation, verification, and adjudication should be separable across actors to reduce conflicts of interest and single points of trust. \\
\addlinespace
Accountability for delegation and proxying
& Control context must be first-class: provenance should express when presence is direct, mediated, compromised, coerced, or remotely operated, enabling responsibility attribution under dispute. \\
\addlinespace
Resilience to laundering, substitution, and manipulation
& Provenance should assume forged paths and compromised sources, supporting corroboration across artefacts and detectability of substitution, replay, or manipulation points. \\
\addlinespace
Portability across borders and sectors
& Evidence semantics and assurance interpretation should remain valid across jurisdictions and institutional boundaries, while allowing sector-specific thresholds and legal tests. \\
\addlinespace
Governance-grade auditability
& Oversight actors must be able to access, interpret, and audit location-linked evidence under safeguards, including clear retention semantics and verification interfaces. \\
\bottomrule
\end{tabularx}
\caption{Summary of design principles for a next-generation digital trust infrastructure for location-data provenance.}
\label{tab:principles_summary}
\end{table}

\hspace{1cm}

These principles have a specific scope and an extensible logic. Their scope comes from the recurring failure modes of location-linked evidence synthesised in Table~\ref{tab:taxonomy}. Their extensible logic comes from the underlying integrity structure, which can also apply to other contested contextual claims, such as authorship, affiliation, or the provenance of digital artefacts more broadly. In later sections, we treat these principles as a stable baseline for sector-specific assurance profiles. They can be extended by adding domain-specific constraints, for instance stronger freshness guarantees in crisis response or stronger chain-of-custody semantics in trade compliance, while preserving the core requirement that location-data provenance remains contestable, privacy-compatible, and usable for adjudication.

The relationship between principles and capabilities is many-to-many. The minimum capabilities in Section~\ref{sec:requirements_summary} are derived primarily from the failure modes in Table~\ref{tab:taxonomy}, while the design principles specify the conditions under which those capabilities can support legitimate contestability. For example, claim-evidence binding responds to fabricated locality, but it must also satisfy proportionality, portability, and auditability. Temporal binding responds to replay and recontextualisation, while context scoping and purpose-bound disclosure determine how that capability remains rights-compatible. Accountable control context responds to proxying and delegated control, while role separation and governance-grade auditability determine how responsibility attribution can be assessed by competent actors. In this sense, the principles operate as cross-cutting constraints on the capabilities developed below.

\subsection{Requirements summary}
\label{sec:requirements_summary}

The seven design principles above are stated at the level of governance objectives and architectural constraints. This subsection translates them into a minimal set of infrastructure capabilities for location-linked claims and evidence. The derivation rule is as follows: a capability is included when it is required to prevent one or more failure modes in Table~\ref{tab:taxonomy} from persisting under otherwise well-intentioned implementations. The requirements below therefore establish necessary conditions for contestability under dispute. Truth, admissibility, and institutional reliance also depend on sector-specific assurance thresholds, legal procedures, and evidentiary standards.

Table~\ref{tab:reqsnapshot} maps each failure mode to a corresponding minimum capability that enables a verifier to challenge the claim through defined evidence and procedures. The mapping is phrased in sector-neutral terms, so that different domains can instantiate these capabilities with different assurance thresholds, legal bases, and institutional workflows.

\begin{table}[b!ht]
\centering
\small
\renewcommand{\arraystretch}{1.12}
\setlength{\tabcolsep}{5.5pt}
\begin{tabularx}{\linewidth}{
    >{\RaggedRight\arraybackslash}p{0.33\linewidth}
    >{\RaggedRight\arraybackslash}X
}
\toprule
\textbf{Failure mode (Table~\ref{tab:taxonomy})} & \textbf{Minimum infrastructure capability for contestability} \\
\midrule
Fabricated locality
& \textbf{Claim-evidence binding:} proof objects that bind the asserted location claim to supporting artefacts and issuance context, reducing the feasibility of purely synthetic local narratives. \\
\addlinespace
Replay and recontextualisation
& \textbf{Temporal binding and context scoping:} verifiable time window, freshness semantics, and claim scope so that artefacts reused outside their valid context become detectable. \\
\addlinespace
Proxying and delegated control
& \textbf{Accountable control context:} expressible and auditable delegation, mediation, compromise, coercion, or remote-operation semantics sufficient to attribute responsibility when presence is simulated through intermediaries. \\
\addlinespace
Origin laundering and documentary substitution
& \textbf{Continuity across intermediaries:} chain-of-custody reasoning and substitution-point detection so that relabelling or documentary replacement becomes detectable under audit. \\
\addlinespace
Signal manipulation and interference
& \textbf{Multi-source robustness:} corroboration across heterogeneous evidence sources and explicit uncertainty semantics to reduce single-source dependence under spoofing, jamming, or telemetry-manipulation conditions. \\
\addlinespace
Record opacity and asymmetric access
& \textbf{Governance-grade audit interfaces:} enforceable access, retention, and verification interfaces for legitimate scrutiny, including privacy-compatible modalities such as controlled environments where appropriate. \\
\addlinespace
Jurisdiction ambiguity and inference overreach
& \textbf{Shared semantics and assurance interpretation:} portable evidence semantics and declared assumptions so that verifiers can avoid over-claiming and can apply sector-specific thresholds consistently. \\
\bottomrule
\end{tabularx}
\caption{Minimum infrastructure capabilities implied by the provenance failure modes in Table~\ref{tab:taxonomy}.}
\label{tab:reqsnapshot}
\end{table}

Two cross-cutting constraints apply to all requirements. First, \textbf{proportionality} must be treated as a design constraint rather than a policy afterthought. Because location is sensitive, contestability should be achieved through selective, purpose-bound disclosure and legally defined access conditions. Second, \textbf{portability} must be preserved. The same evidence object should remain interpretable across jurisdictions and sectors, even when the legal test, risk tolerance, or enforcement threshold differs. These constraints ensure that location-data provenance strengthens sovereignty and accountability while preserving rights compatibility, institutional legitimacy, and cross-sector interoperability.

The next section uses these requirements to frame the solution space at a conceptual level and introduces proof-of-location as a candidate capability layer for turning location into a digital primitive within a next-generation digital trust infrastructure for location-data provenance.

\section{Location as a digital primitive}
\label{sec:location-primitive}
The preceding sections framed location-data provenance as an evidentiary problem of high-impact digital decisions that increasingly depend on where an event occurred, while location assertions are often treated as low-integrity metadata instead of evidence-grade claims. When location becomes a condition for access, eligibility, liability, attribution, or jurisdiction, it functions as a governance boundary in digital systems. In that setting, coordinates alone are insufficient. What matters is whether a location-linked claim can be justified by verifiable observations and constraints that a third party can evaluate under adversarial pressure. Early security work already recognised the strategic importance of binding digital actions to physical geography as a way to ground otherwise unbounded cyberspace \autocite{denning1996location}.

This motivates treating location as a first-class digital primitive alongside identity and time. A digital primitive denotes a reusable, machine-verifiable claim type with explicit semantics, assurance assumptions, and verification procedures. As identity systems evolved from self-asserted names to verifiable credentials, location-sensitive services increasingly require verifiable presence claims instead of self-reported coordinates. A broader cryptographic line of work formalises what can and cannot be proven from physical position, including models of position-based cryptography and their limitations under collusion \autocite{chandran2009position}. The relevance for this paper is practical: if location is to be used as evidence in governance contexts, then it must be expressible as a claim with explicit assumptions, a bounded time window, and a verification procedure that can survive contestation.

\begin{figure}
    \centering
    \includegraphics[width=0.8\linewidth]{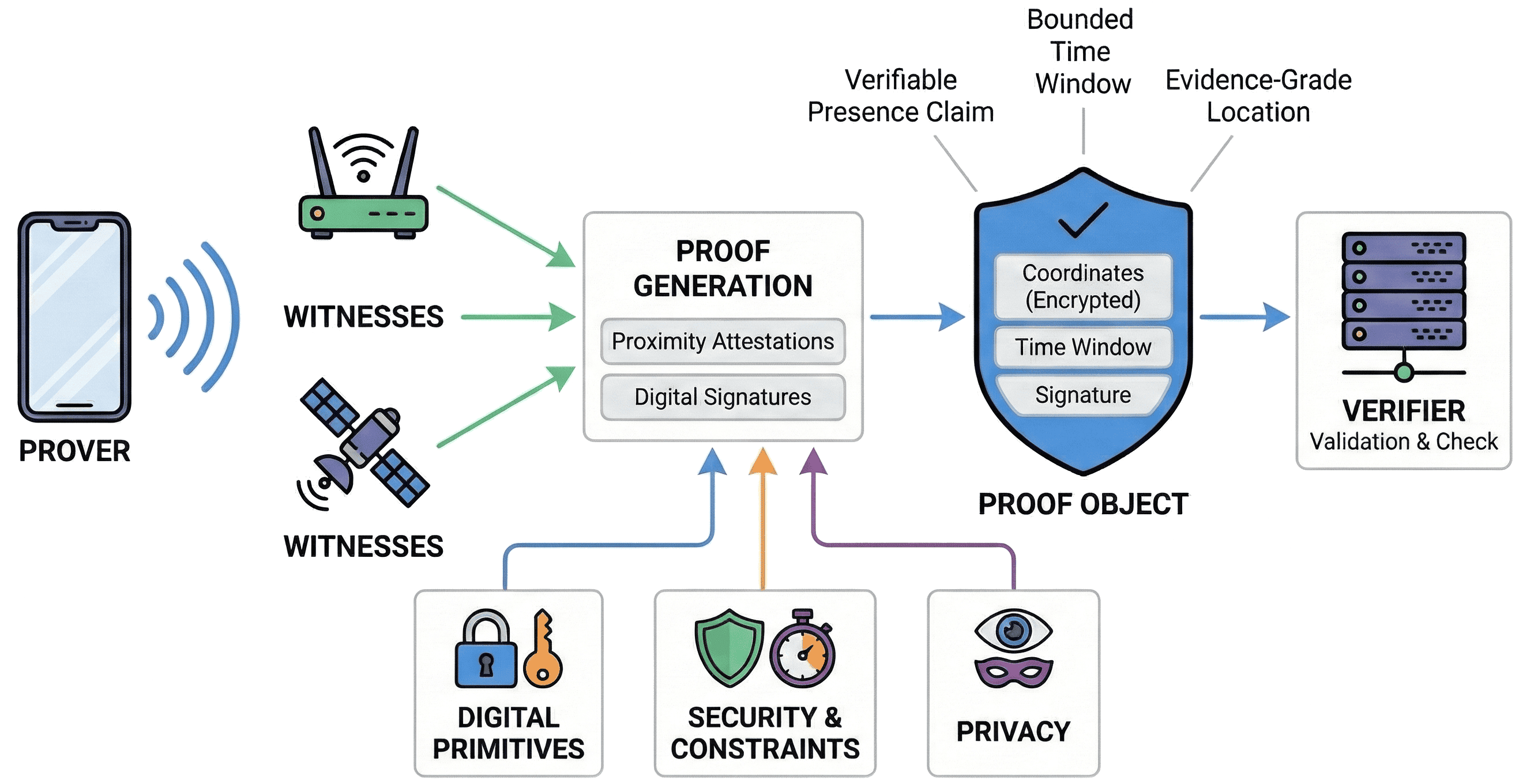}
    \caption{Illustration of a Proof-of-location (PoL) protocol involving a Prover, Witnesses, and a Verifier. The Prover seeks to demonstrate presence at a location, Witnesses provide attestations of proximity, and the Verifier checks the resulting proof object.}
    \label{fig:pol_diagram}
\end{figure}

Proof-of-location (PoL) systems are a concrete family of mechanisms for producing such evidence-grade presence claims. A PoL protocol typically involves a \emph{prover} seeking to demonstrate presence, one or more nearby \emph{witnesses} providing proximity attestations, for instance through digital signatures, and a \emph{verifier} that checks the resulting proof object \autocite{saroiu2009enabling,luo2010proving,brito2025taxonomy}. Figure~\ref{fig:pol_diagram} illustrates this basic architecture. PoL extends localisation and secure positioning. Secure positioning aims to compute coordinates robustly under attack, for example through verifiable multilateration and consistency checks among measurements \autocite{capkun2006secure}. PoL targets verifiable presence as a reusable and portable credential: the verifier may require assurance that the prover was within a specified region during a specified time window, under a declared threat model, while exact coordinates may be unnecessary for the governance decision at hand.

For governance use, a PoL proof object must express more than the fact that a protocol was executed. It should specify the asserted region, the bounded time window, the evidence sources or witness policy, the applicable threat assumptions, the verification procedure, and the disclosure constraints under which the claim can be checked. These elements connect PoL to the minimum capabilities derived in Section~\ref{sec:requirements_summary}: claim-evidence binding, temporal binding, accountable control context, multi-source robustness, and shared assurance interpretation. They also make explicit the distinction between proving proximity to a place and supporting a contestable institutional claim about presence, origin, routing, jurisdiction, or control.

The provenance failure modes in Table~\ref{tab:taxonomy} clarify the adversarial context PoL must address. At minimum, protocols must anticipate spoofing of software-reported location, relay and proxy attacks that simulate proximity, and collusion among parties who jointly fabricate a claim. Distance-bounding protocols were introduced to constrain relay-style fraud by enforcing physical limits on signal round-trip time, and have since been studied as proximity primitives with well-understood security and implementation trade-offs \autocite{brands1993distance,avoine2018security}. In parallel, privacy is a core requirement. Location proofs can easily become surveillance artefacts if they are linkable, overly precise, or over-collected. A substantial line of work therefore aims to make PoL both verifiable and privacy-preserving, for example by limiting disclosure and addressing issuer-side manipulations such as geo-tampering attacks that silently undermine proof semantics \autocite{luo2010proving,akand2021privacy,bogdanov2025zero}.

\paragraph{Capability alignment.}
In the logic of this paper, PoL is introduced because it gives a concrete protocol-level interpretation to five of the minimum capabilities derived in Section~\ref{sec:requirements_summary}. First, it supports \textbf{claim-evidence binding} by linking a presence claim to witness attestations, verification metadata, and a protocol execution context \autocite{saroiu2009enabling,luo2010proving,brito2025taxonomy}. Second, it supports \textbf{temporal binding and context scoping} when the proof object carries a bounded time window, freshness assumptions, and a declared region or proximity policy; this is especially relevant for capture-time authenticity and content-provenance settings, where the evidentiary value of a media object depends on whether the claimed place and time of capture can be bound to the artefact \autocite{brito2026capturetime_pol}. Third, it supports \textbf{multi-source robustness} when attestations are produced by multiple independent or policy-qualified witnesses, reducing dependence on a single reported coordinate, signal source, or authority \autocite{Brito2025SciReports}. Fourth, it can support \textbf{accountable control context} when the proof semantics distinguish direct presence, mediated presence, delegated operation, or proxied interaction, although responsibility attribution also requires institutional interpretation. Fifth, it supports \textbf{shared semantics and assurance interpretation} when the proof object declares its region semantics, witness policy, threat assumptions, verification procedure, and relationship to adjacent provenance systems such as content credentials \autocite{c2pa2024spec,brito2026capturetime_pol}. Secure positioning and distance-bounding work clarify the physical and timing assumptions relevant to spoofing and relay resistance \autocite{capkun2006secure,brands1993distance,avoine2018security}, while privacy-preserving and zero-knowledge PoL variants show how verifiability can be combined with selective disclosure \autocite{akand2021privacy,bogdanov2025zero}.

Existing PoL literature provides mechanisms for producing and checking presence proofs, however several legal-technical questions remain underdeveloped. These include evidentiary admissibility, burden of proof, assurance levels, retention duties, audit rights, cross-border recognition, institutional challenge procedures, chain-of-custody across organisations, interpretation of delegated or coerced control, and the relationship between location proofs and sector-specific legal thresholds. These gaps explain why PoL is treated here as an \emph{evidence-generation capability layer} within a broader location-data provenance framework. Contestability under dispute depends on both verifiable proof objects and institutional procedures for interpreting, challenging, and relying on them.

Therefore, PoL is most useful as a facilitator concept for a next-generation digital trust infrastructure because it can turn selected location assertions into verifiable presence claims that are portable across domains and composable with digital identity systems, content provenance frameworks, and governance-grade audit procedures. Decentralised PoL architectures further reduce dependence on a single location authority by distributing attestation across sets of entities, so that trust emerges from collective agreement and auditable evidence production, adding territorial sovereignty over location data as a design goal \autocite{Brito2025SciReports}. This motivates why location should be treated as a digital primitive and positions PoL as a candidate mechanism for implementing evidence-grade location-data provenance in future governance contexts.

\section{Conclusion} 
\label{sec:conclusion}
This paper argued that location-data provenance has become a strategic requirement for European digital sovereignty and security. Across elections and the information environment, cross-border supply chains, financial and sanctions compliance, and safety-relevant infrastructure, institutions increasingly rely on location-linked claims to allocate responsibility, trigger legal regimes, and justify high-impact decisions. Yet such claims are still treated as low-integrity metadata, resulting in recurring trust erosion where fabricated locality, replay, proxying, laundering, and record opacity allow adversaries to operate at scale while institutions remain dependent on slow reconstruction or opaque intermediaries.

To address this gap, the paper introduced a cross-sector risk taxonomy that decomposes provenance failures into integrity axes and recurring failure modes, and derived design principles for next-generation digital trust infrastructure. The central objective is contestability under dispute: the ability to challenge and adjudicate location-linked claims in a structured, privacy-compatible manner under adversarial pressure. Framed as a digital primitive, location must be expressible as evidence-bearing claims rather than self-asserted coordinates. Proof-of-Location provides one family of mechanisms for producing verifiable presence claims under explicit threat and privacy assumptions, including decentralised approaches that reduce reliance on single authorities.

Looking ahead, the core challenge is to standardise how location-linked claims are expressed, verified, and contested as evidence. Future work should focus on shared evidence semantics, assurance levels, and interoperability profiles that enable consistent evaluation across sectors and jurisdictions. In parallel, institutional integration into supervisory and enforcement workflows should preserve accountability and fundamental rights without creating surveillance-by-default.

\section*{Acknowledgments}
This work was supported in part by the 2025/2026 Charlemagne Prize Academy Fellowship of the Charlemagne Prize Foundation.

\printbibliography

\end{document}